\documentclass[a4paper,epjc3]{svjour3}
\pdfoutput=1
\usepackage{amsmath,amssymb}
\usepackage{color,xcolor}
\definecolor{blue}{rgb}{0,0,0.5}
\usepackage[colorlinks,linkcolor=blue,urlcolor=blue,citecolor=blue]{hyperref}
\usepackage{graphicx}
\usepackage{cite}
\usepackage{xspace}
\usepackage{relsize}
\usepackage{bm}
\usepackage{multirow}
\usepackage{booktabs}
\usepackage{enumerate}
\usepackage{lmodern}
\usepackage[utf8]{inputenc}

\allowdisplaybreaks
\definecolor{maroon}{cmyk}{0, 0.87, 0.68, 0.32}
\definecolor{halfgray}{gray}{0.55}
\definecolor{ipython_frame}{RGB}{207, 207, 207}
\definecolor{ipython_bg}{RGB}{247, 247, 247}
\definecolor{ipython_red}{RGB}{186, 33, 33}
\definecolor{ipython_green}{RGB}{0, 128, 0}
\definecolor{ipython_cyan}{RGB}{64, 128, 128}
\definecolor{ipython_purple}{RGB}{170, 34, 255}

\usepackage{listings}
\lstdefinelanguage{iPython}{
    morekeywords={access,and,break,class,continue,def,del,elif,else,except,exec,finally,for,from,global,if,import,in,is,lambda,not,or,pass,print,raise,return,try,while},%
    morekeywords=[2]{abs,all,any,basestring,bin,bool,bytearray,callable,chr,classmethod,cmp,compile,complex,delattr,dict,dir,divmod,enumerate,eval,execfile,file,filter,float,format,frozenset,getattr,globals,hasattr,hash,help,hex,id,input,int,isinstance,issubclass,iter,len,list,locals,long,map,max,memoryview,min,next,object,oct,open,ord,pow,property,range,raw_input,reduce,reload,repr,reversed,round,set,setattr,slice,sorted,staticmethod,str,sum,super,tuple,type,unichr,unicode,vars,xrange,zip,apply,buffer,coerce,intern},%
    sensitive=true,%
    morecomment=[l]\#,%
    morestring=[b]',%
    morestring=[b]",%
    morestring=[s]{'''}{'''},
    morestring=[s]{"""}{"""},
    morestring=[s]{r'}{'},
    morestring=[s]{r"}{"},%
    morestring=[s]{r'''}{'''},%
    morestring=[s]{r"""}{"""},%
    morestring=[s]{u'}{'},
    morestring=[s]{u"}{"},%
    morestring=[s]{u'''}{'''},%
    morestring=[s]{u"""}{"""},%
    literate=
    {á}{{\'a}}1 {é}{{\'e}}1 {í}{{\'i}}1 {ó}{{\'o}}1 {ú}{{\'u}}1
    {Á}{{\'A}}1 {É}{{\'E}}1 {Í}{{\'I}}1 {Ó}{{\'O}}1 {Ú}{{\'U}}1
    {à}{{\`a}}1 {è}{{\`e}}1 {ì}{{\`i}}1 {ò}{{\`o}}1 {ù}{{\`u}}1
    {À}{{\`A}}1 {È}{{\'E}}1 {Ì}{{\`I}}1 {Ò}{{\`O}}1 {Ù}{{\`U}}1
    {ä}{{\"a}}1 {ë}{{\"e}}1 {ï}{{\"i}}1 {ö}{{\"o}}1 {ü}{{\"u}}1
    {Ä}{{\"A}}1 {Ë}{{\"E}}1 {Ï}{{\"I}}1 {Ö}{{\"O}}1 {Ü}{{\"U}}1
    {â}{{\^a}}1 {ê}{{\^e}}1 {î}{{\^i}}1 {ô}{{\^o}}1 {û}{{\^u}}1
    {Â}{{\^A}}1 {Ê}{{\^E}}1 {Î}{{\^I}}1 {Ô}{{\^O}}1 {Û}{{\^U}}1
    {œ}{{\oe}}1 {Œ}{{\OE}}1 {æ}{{\ae}}1 {Æ}{{\AE}}1 {ß}{{\ss}}1
    {ç}{{\c c}}1 {Ç}{{\c C}}1 {ø}{{\o}}1 {å}{{\r a}}1 {Å}{{\r A}}1
    {€}{{\EUR}}1 {£}{{\pounds}}1
    {^}{{{\color{ipython_purple}\^{}}}}1
    {=}{{{\color{ipython_purple}=}}}1
    {+}{{{\color{ipython_purple}+}}}1
    {*}{{{\color{ipython_purple}$^\ast$}}}1
    {/}{{{\color{ipython_purple}/}}}1
    {+=}{{{+=}}}1
    {-=}{{{-=}}}1
    {*=}{{{$^\ast$=}}}1
    {/=}{{{/=}}}1,
    literate=
    *{-}{{{\color{ipython_purple}-}}}1
     {?}{{{\color{ipython_purple}?}}}1,
    identifierstyle=\color{black}\ttfamily,
    commentstyle=\color{ipython_cyan}\ttfamily,
    stringstyle=\color{ipython_red}\ttfamily,
    keepspaces=true,
    showspaces=false,
    showstringspaces=false,
    rulecolor=\color{ipython_frame},
    frame=single,
    frameround={t}{t}{t}{t},
    framexleftmargin=6mm,
    numbers=left,
    numberstyle=\tiny\color{halfgray},
    backgroundcolor=\color{ipython_bg},
    basicstyle=\footnotesize\ttfamily,
    keywordstyle=\color{ipython_green}\ttfamily,
    aboveskip=1.2em,
    belowskip=1.2em,
}

\journalname{Eur. Phys. J. C}
\begin{document}

\title{\includegraphics[width=4cm]{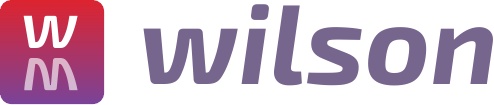}}
\subtitle{A Python package for the running and matching of Wilson coefficients above and below the electroweak scale}
\titlerunning{\texttt{wilson}}

\author{Jason Aebischer\thanksref{e1,addr1}
\and
Jacky Kumar\thanksref{e2,addr2}
        \and
        David~M.~Straub\thanksref{e3,addr1}
}

\thankstext{e1}{e-mail: jason.aebischer@tum.de}
\thankstext{e2}{e-mail: jkumar@iisermohali.ac.in}
\thankstext{e3}{e-mail: david.straub@tum.de}

\institute{Excellence Cluster Universe, TUM, Boltzmannstr.~2, 85748~Garching, Germany \label{addr1}
\and
Department of Physics, Indian Institute of Science Education and Research, \\
Mohali, Punjab, 140036 India  \label{addr2}
}

\date{}

\newcommand{\wilson}{\textsf{\emph{wilson}}}
\newcommand{\msbar}{$\overline{\text{MS}}$}

\maketitle

\begin{abstract}

\wilson\ is a Python library for matching and running Wilson coefficients of higher-dimensional operators beyond the Standard Model. Provided with the numerical values of the Wilson coefficients at a high new physics scale, it automatically performs the renormalization group evolution within the Standard Model effective field theory (SMEFT), matching onto the weak effective theory (WET) at the electroweak scale, and QCD/QED renormalization group evolution below the electroweak scale down to hadronic scales relevant for low-energy precision tests. The matching and running encompasses the complete set of dimension-six operators in both SMEFT and WET.
The program builds on the Wilson coefficient exchange format (WCxf) and can thus be easily combined with a number of existing public codes.
\end{abstract}

\section{Introduction}

The Standard Model (SM) \cite{Weinberg:1967tq, Glashow:1961tr, Salam:1968rm} is considered to be an effective theory valid only up to a new physics scale $\Lambda$, which negative searches for new particles at the LHC likely relegate to well above the electroweak (EW) scale.
If no light degrees of freedom beyond the SM are assumed, any new physics effect in processes proceeding at energies well below $\Lambda$ can be described by local interactions among SM fields invariant under the SM gauge symmetry \cite{Buchmuller:1985jz, Grzadkowski:2010es}.
This effective field theory (EFT) approach \cite{Appelquist:1974tg, Wilson:1993dy} to new physics not only allows to resum large logarithms that might invalidate calculations in perturbation theory for vastly different scales relevant in a given process, but also serves as a convenient intermediate step between ``model building'' in the UV and low-energy phenomenology. If new physics predictions for experimental observables are expressed in terms of Wilson coefficients of an EFT beyond the SM, the investigation of the low-energy implications of a concrete new physics model becomes much simpler since only the Wilson coefficients need to be calculated at the appropriate scale.

While the EFT approach to new physics has been ubiquitous in quark flavour physics -- dealing with processes at energies of few GeV -- for a long time already, the experimental indications that $\Lambda$ lies well above the electroweak scale have led to the realization that this approach is also valuable for processes of electroweak scale energies like Higgs physics or electroweak precision tests (see \cite{Ellis:2018gqa} and references therein). In contrast to the EFT \textit{below} the electroweak scale, that is conventionally called the weak effective theory (WET) \cite{Jenkins:2017jig, Aebischer:2017gaw, Jenkins:2017dyc} and only contains QED and QCD gauge interactions, the EFT \textit{above} the electroweak scale, conventionally called SMEFT\footnote{Throughout, we work with the EFT above the electroweak scale with \textit{linearly} realized electroweak symmetry breaking (see \cite{deFlorian:2016spz} and references therein for a discussion of the non-linear case).}\cite{Weinberg:1980wa, Coleman:1969sm, Callan:1969sn},  contains $SU(2)_L$ interactions that do not conserve flavour. Consequently, quantum effects lead to an interesting interplay between processes with and without flavour change and call for a \textit{global} approach.

Starting from the new physics scale $\Lambda$, the phenomenological analysis of a UV model typically requires the following technical steps.\footnote{Steps 3.-5.\ can be omitted for observables at electroweak scale energies.}
\begin{enumerate}
  \item Compute the SMEFT Wilson coefficients at $\Lambda$.
  \item Perform the renormalization group (RG) evolution of the SMEFT Wilson coefficients down to the electroweak scale.
  \item Match the complete set of SMEFT Wilson coefficients onto the WET.
  \item Perform the RG evolution of WET Wilson coefficients.
  \item If the process proceeds at energies below the $b$ quark mass, repeat the last two steps for the WET with reduced numbers of quark and lepton flavours as appropriate.
  \item Compute the process of interest as a function of the low-energy Wilson coefficients.
\end{enumerate}
While the first five steps are straightforward in principle, the full procedure is technically challenging in practice due to the vast number of Wilson coefficients already at dimension six (cf. \cite{Alonso:2013hga,Jenkins:2017jig}).
The \wilson\ package provides an automated solution to steps 2.-5.\ above.
Given the SMEFT Wilson coefficients at the UV scale $\Lambda$, it bridges the gap to the low-energy phenomenology in step 6., which is implemented in other public codes such as \texttt{flavio} \cite{Straub:2018kue}.
The package makes use of the following results in the literature.
\begin{itemize}
  \item The complete basis of SMEFT operators first derived in \cite{Buchmuller:1985jz} and for a non-redundant set of operators in \cite{Grzadkowski:2010es}.
  \item The complete one-loop RG evolution in SMEFT \cite{Jenkins:2013zja,Jenkins:2013wua,Alonso:2013hga}.
  \item Analytical solutions to the one-loop RG evolution of all flavour violating operators in WET \cite{Aebischer:2017gaw}.
  \item The complete RG evolution of WET operators \cite{Jenkins:2017dyc}.
  \item The complete tree-level matching of SMEFT onto the WET \cite{Aebischer:2015fzz,Jenkins:2017jig}.
  \item The definition of a Wilson coefficient exchange format (WCxf) that allows to define EFTs, bases of Wilson coefficients, and facilitates exchanging numerical values of Wilson coefficients between different codes \cite{Aebischer:2017ugx}.
\end{itemize}
It benefits from the following public physics codes:
\begin{itemize}
  \item The SMEFT RG evolution was ported from (and is tested against) the DsixTools Mathematica package \cite{Celis:2017hod}.
  \item The QCD evolution of quark masses and the strong coupling constant is computed with the \texttt{python-rundec} package that wraps the \texttt{CRunDec} module \cite{Herren:2017osy}.
  \item The SM $\overline{\text{MS}}$ parameters at the electroweak scale have been obtained with the \texttt{mr} package \cite{Kniehl:2016enc}.
\end{itemize}

The rest of this note is organized as follows.
In section~\ref{sec:desc}, we give some details on the implementation of running and matching in \wilson.
Section~\ref{sec:inst} describes how to install the package.
Section~\ref{sec:usage} contains details on how to use the code.
In section~\ref{sec:phys}, we present a simple example application, reproducing a well-known result from the literature.

\section{Description}\label{sec:desc}

\begin{figure}
  \includegraphics[trim={3cm 3cm 3cm 3cm},clip,width=\textwidth]{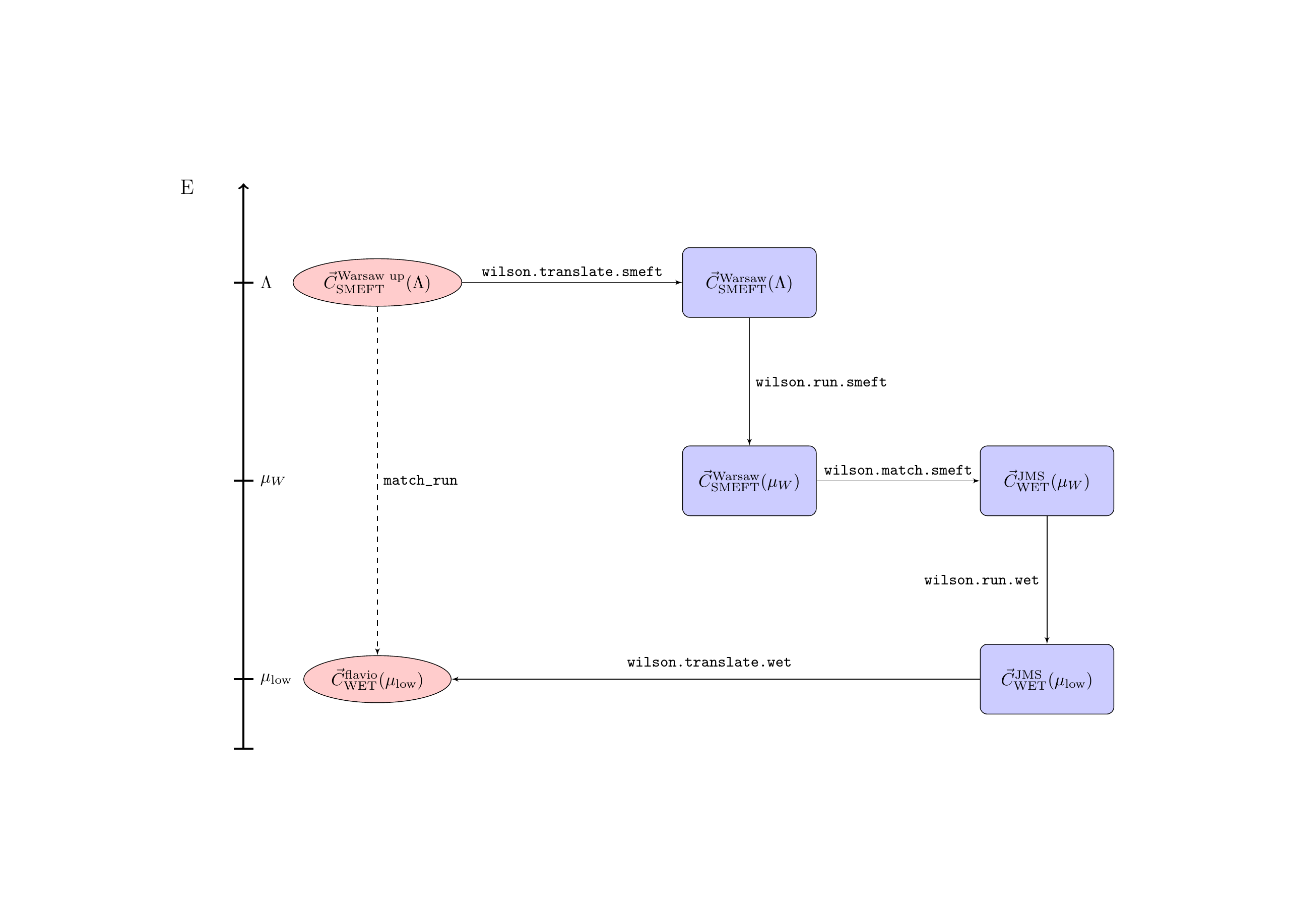}
  \caption{Typical internal workflow in the \wilson\ package: Starting from the SMEFT Wilson coefficients at the scale $\Lambda$, various submodules take care of the necessary basis translations, RG running, and matching to finally obtain the WET Wilson coefficients at the low scale. From a user perspective, the \texttt{match\char`_run} method (see section~\ref{sec:matchrun}) performs all these steps automatically.}
  \label{fig:flow}
\end{figure}

The \wilson\ package consists of several submodules taking care of the RG evolution, basis translation, and matching. A typical internal workflow is shown in figure~\ref{fig:flow}, where a set of SMEFT Wilson coefficients in the ``Warsaw up'' basis \cite{Jenkins:2017jig} at the scale $\Lambda$ is the input and the WET Wilson coefficients at the scale $\mu_\text{low}$ are returned in the basis used by the \texttt{flavio} package. Internally, the Warsaw basis as defined in WCxf \cite{Aebischer:2017ugx} is used for the SMEFT running, and the JMS basis \cite{Jenkins:2017jig} for the matching and WET running.
From a user's perspective, the entire procedure is performed automatically when using the \texttt{match\char`_run} method described in section~\ref{sec:matchrun}, as indicated by the dashed arrow.
Below, we discuss some implementation details of the individual submodules.

\subsection{Extraction of Standard Model parameters in SMEFT}

Starting from a set of Wilson coefficients at the UV scale, given e.g.\ in WCxf format, to solve the SMEFT RGEs one additionally requires the values of SM parameters like gauge couplings, Yukawa couplings, and Higgs potential parameters. This is challenging for two reasons. First, these parameters are experimentally determined at the electroweak scale or below, and their evolution to the UV scale depends on the SMEFT Wilson coefficients themselves. Second, the experimental extraction itself is subject to dimension six corrections already at tree level. To solve these two problems, we proceed in three steps.
\begin{enumerate}
  \item We determine all the SM parameters in the \msbar\ scheme \cite{Bardeen:1978yd} at the scale $M_Z$.
  \item We invert the relations between the effective \msbar\ SM parameters and their counterparts in SMEFT, that are given e.g.\ in \cite{Dedes:2017zog}.
  \item We iteratively determine the SM parameters at the UV scale by running up and down with the SM boundary conditions imposed at the scale $M_Z$ and the Wilson coefficient boundary conditions at the UV scale $\Lambda$.
\end{enumerate}

Concerning the first step, the SM \msbar\ parameters used by us are listed in table~\ref{tab:smpar}.
The following comments are in order.
\begin{itemize}
  \item For the running of the quark masses to the scale $M_Z$, we have used the \texttt{python-rundec} package \cite{Herren:2017osy}.
  \item For the determination of the running top, $W$, $Z$, and Higgs masses, we have used the \texttt{mr} package \cite{Kniehl:2016enc}.
  \item For the lepton masses, we have neglected the $O(\alpha_e)$ shift from the conversion to the \msbar\ scheme.
  \item We do not display uncertainties as fixed values are used in the code. We expect the parametric errors to be subdominant to other uncertainties in the calculation, e.g.\ from the iterative determination of high-scale SM parameters.\footnote{%
To check the accuracy of the iterative determination of SM parameters, the class \texttt{wilson.run.smeft.SMEFT}, that is initialized by a \texttt{wcxf.WC} instance, provides a method
\texttt{get\char`_smpar}, that computes the predicted values for the SM \msbar\ parameters at the electroweak scale, which should correspond to the values in table~\ref{tab:smpar}.}
\end{itemize}

\begin{table}
\centering
\begin{tabular}{llll}
\toprule
par. & value & par. & value \\
\midrule
$\alpha_e$ & $1 / 127.9$ & $m_u$ & $0.00127$ \\
$\alpha_s$ & $0.1185$ & $m_d$ & $0.00270$ \\
$V_{us}$ &  $0.2243$ & $m_s$ & $0.0551$ \\
$V_{cb}$ & $0.04221$ & $m_c$ & $0.635$ \\
$V_{ub}$ & $0.00362$ & $m_b$ & $2.85$ \\
$\gamma$ & $1.27$ & $m_t$ & $169.0$ \\
$m_e$ & $0.000511$ & $m_W$ & $80.20$ \\
$m_{\mu}$ & $0.1057$ & $m_Z$ & $91.46$ \\
$m_{\tau}$ & $1.777$ & $m_h$ & $130.6$ \\
\bottomrule
\end{tabular}
\caption{SM \msbar\ parameters at the scale $M_Z$. Masses are given in units of GeV.}
\label{tab:smpar}
\end{table}

We note that we treat the CKM elements as elements of a unitary $3\times 3$ matrix. Dimension-six contributions to the $W$ coupling to quarks are thus \emph{not} absorbed in effective CKM elements, as done e.g.\ in \cite{Dedes:2017zog}. We find this procedure more convenient for our purposes; in particular, it allows to continue to use unitarity relations in low-energy calculations in flavour physics. While this blurs the connection between these CKM elements and the semi-leptonic decays that are used to measure them, we note that this connection is anyway blurred in SMEFT due to direct dimension-six four-fermion contributions to these decays that can lead to a process-dependent shift of the apparent CKM element (see e.g.\ \cite{Gonzalez-Alonso:2016etj} for a discussion of $s\to u$ transitions and \cite{Jung:2018lfu} for $b\to c$ transitions).

\subsection{RG evolution in SMEFT}

Once the SM parameters at the input scale have been determined, the SMEFT RGEs, that have the form
\begin{equation}
  \frac{d C_i}{d\ln\mu} = \frac{1}{16\pi^2} \sum_j \gamma_{ji} C_i \,,
\end{equation}
can be solved numerically by integrating the right-hand side.
Our implementation closely follows the DsixTools package \cite{Celis:2017hod}.

As an important caveat, we caution the reader that the numerical
inputs and outputs, using the non-redundant basis defined by
the WCxf convention, differ from the conventions used in
\cite{Jenkins:2013zja,Jenkins:2013wua,Alonso:2013hga},
where a redundant basis of flavour indices is employed,
by symmetry factors in some cases. We refer to appendix~A of \cite{Aebischer:2018iyb}, where this issue is discussed in detail.

\subsection{Matching from SMEFT to WET}

We implement the complete tree-level matching from SMEFT to WET as derived in \cite{Jenkins:2017jig}. It includes the full set of non-redundant gauge-invariant dimension six operators in both theories. The matching is performed at the EW scale.

\subsection{RG evolution in WET}

In the weak effective theory, the dimension-6 operators are renormalized by QCD and QED. Analytical solutions to the one-loop RGEs of all quark flavour violating operators have been presented in \cite{Aebischer:2017gaw}\footnote{See also \cite{Buchalla:1995vs,Buras:2011we} and references therein.}. To extend this to the \emph{complete} operator basis\footnote{In the WET RG evolution, we restrict ourselves to baryon and lepton number conserving operators for the time being.} of WET, we proceed in three steps.
\begin{enumerate}
  \item We take the beta functions from \cite{Jenkins:2017jig}, discarding terms that are quadratic in dipole operator coefficients (these terms correspond to dimension eight contributions when matching from the SMEFT with linearly realized electroweak symmetry breaking).
  \item We rescale dipole operators and three-gluon operators in the following way:
  \begin{align}
    \bar f^i_{L} \sigma^{\mu \nu} f^j_{R}\, F_{\mu \nu}
    &\to \frac{e}{g_s^2}m_f\bar f^i_{L} \sigma^{\mu \nu} f^j_{R}\, F_{\mu \nu} \,,
    \\
    \bar f^i_{L} \sigma^{\mu \nu} T^A f^j_{R}\, G^A_{\mu \nu}
    &\to \frac{1}{g_s} m_f \bar f^i_{L} \sigma^{\mu \nu} T^A f^j_{R}\, G^A_{\mu \nu} \,,
    \\
     G_\mu^{A\nu} G_\nu^{B\rho} G_\rho^{C\mu}
     &\to \frac{1}{g_s}G_\mu^{A\nu} G_\nu^{B\rho} G_\rho^{C\mu} \,,
  \end{align}
  where $m_f=\text{max}(m_{f_i}, m_{f_j})$.
  This allows us to write the RGEs in the simple form
  \begin{equation}
    \frac{d C_i}{d\ln\mu} = \frac{g_s^2}{16\pi^2} \sum_j \gamma^s_{ji} C_i+ \frac{e^2}{16\pi^2} \sum_j \gamma^e_{ji} C_i \,.
    \label{eq:wetrge}
  \end{equation}
  Note in particular that there are no linear or mixed terms in $g_s$ or $e$.
  Thanks to the rescalings, the anomalous dimension matrices $\gamma^{s,e}$ only
  contain numbers and ratios of fermion masses, which are RG invariant to $O(\alpha_s)$ and thus can be treated as constants to good approximation.
  \item Having rewritten the RGEs in the simple form \eqref{eq:wetrge}, we can use the procedure described in \cite{Aebischer:2017gaw} to obtain the QCD and QED evolution matrices that solve the RGE as
  \begin{equation}
    C_i(\mu) = \left[U_s(\mu, \mu_0)_{ij} + \Delta U_e(\mu, \mu_0)_{ij}\right]C_j(\mu_0) \,.
  \end{equation}
\end{enumerate}

\section{Installation}\label{sec:inst}

Installing \wilson\ only requires a system with Python version 3.5 or above. It works on Linux, Mac~OS, and Windows.
The most recent version can be installed directly from the Python package index by issuing the command\footnote{The name of the Python 3 executable might differ depending on the system.}
\begin{lstlisting}[language=iPython]
python3 -m pip install wilson --user
\end{lstlisting}
in the terminal, without root privileges.
This will automatically install the \texttt{wcxf} package and command line interface as well, if not already available on the system.
When a new version is available, the package can be upgraded with
\begin{lstlisting}[language=iPython]
python3 -m pip install --upgrade wilson --user
\end{lstlisting}

\section{Usage}\label{sec:usage}

\subsection{Initializiation}\label{sec:init}

Using the \wilson\ package in a Python script or interactive session starts by creating a \texttt{Wilson} object that represents a point in EFT parameter space. On creating the instance, initial values of the Wilson coefficients have to be specified at some scale, e.g.\ the new physics scale $\Lambda$, in a given EFT and basis. For example, the commands
\begin{lstlisting}[language=iPython]
from wilson import Wilson
mywilson = Wilson({'uG_33': 1e-6},
                  scale=1e3, eft='SMEFT', basis='Warsaw')
\end{lstlisting}
create a new \texttt{Wilson} instance where the Wilson coefficient of the chromomagnetic operator with two top quarks in the SMEFT Warsaw basis,
\begin{equation}
O_{uG}^{33} = \left( \bar q_3 \sigma^{\mu \nu} T^A u_3 \right) \widetilde \varphi G_{\mu \nu}^A \,,
\end{equation}
is set to the value $1 /\text{TeV}^2$ at the scale 1~TeV (note that all dimensionful quantities have to be specified in appropriate powers of GeV, as required by WCxf).
At this point, it is important to emphasize the difference between \wilson{}'s \texttt{Wilson} class and the \texttt{WC} class provided by the \texttt{wcxf} Python package:
\begin{itemize}
  \item \texttt{wcxf.WC} represents a set of numerical Wilson coefficients at a \textit{fixed} scale in a \textit{fixed} EFT and basis;
  \item \texttt{wilson.Wilson} represents a point in the parameter space of the EFT beyond the SM, that can be evolved to different scales and translated to different bases within the same EFT without loss of generality, or matched to EFTs valid at lower energies.
\end{itemize}
In fact, after initializing the above object, the Wilson coefficient values at the initial scale can be returned as a \texttt{WC} object simply with \texttt{mywilson.wc}. Likewise, a \texttt{Wilson} object can be easily initialized by loading Wilson coefficient values from a file in WCxf format:
\begin{lstlisting}[language=iPython]
from wilson import Wilson
with open('my_wcxf.json') as f:
    mywilson = Wilson.load_wc(f)
\end{lstlisting}

\subsection{Matching and running}\label{sec:matchrun}

Running, i.e.\ performing the RG evolution in SMEFT and WET, as well as matching from SMEFT to WET (and from WET with five active quark flavours to the variants of WET valid below the bottom and charm mass scales) is the main purpose of the \wilson\ package. Having initialized a \texttt{Wilson} object as described in section~\ref{sec:init} -- we will continue to call this instance \texttt{mywilson} -- the user can obtain Wilson coefficient values (in the form of \texttt{wcxf.WC} instances) in different EFTs, at different scales, in different bases, through the method \verb+match_run+\footnote{Tiny non-zero entries of Wilson coefficients can be traced back to a finite numerical precision used for the solution of the RGEs and can safely be neglected.}:
\begin{lstlisting}[language=iPython]
# running up to 100 TeV and translating to the 'Warsaw up' basis
wc = mywilson.match_run(scale=1e5, eft='SMEFT', basis='Warsaw up')
# running and matching to WET at 100 GeV in the 'JMS' basis
wc = mywilson.match_run(scale=100, eft='WET', basis='JMS')
# running and matching to WET-3 at 2 GeV in the 'flavio' basis
wc = mywilson.match_run(scale=2, eft='WET-3', basis='flavio')
\end{lstlisting}
The names of admissible EFTs and bases can be found on the WCxf website \cite{wcxf-web}.

We note that the output scale can also be higher than the input scale,
but only if the output EFT is the same as the input EFT. In this case,
the RG evolution (in WET or SMEFT) will be performed from the low
input scale to the high output scale. Since the matching is not bijective,
this cannot be done across EFT thresholds.

The default behaviour of the \texttt{Wilson} class can be modified
with a few user options that can be modified either on a single
instance or globally for all future instances of the class
(e.g.\ when importing the package),
\begin{lstlisting}[language=iPython]
mywilson.set_option(OPTION, VALUE)        # set option on instance
Wilson.set_default_option(OPTION, VALUE)  # set option globally
\end{lstlisting}
The following options are implemented as of version~1.4,
\begin{itemize}
  \item \texttt{'smeft\char`_accuracy'} -- set accuracy of the SMEFT RG evolution to numerical integration (value \texttt{'integrate'}, default) or leading-logarithmic approximation (\texttt{'leadinglog'}), which is less accurate but much faster.
  \item \texttt{'qcd\char`_order'}, \texttt{'qed\char`_order'} -- set the order of QED and QCD anomalous dimensions to be taken into account in the WET RG running. Currently both values are restricted to 1 (default, leading order) or 0 (off).
  \item \texttt{'smeft\char`_matchingscale'} -- set the scale (in GeV) where SMEFT is matched onto WET. Defaults to 91.1876 (the central value of the $Z^0$ mass).
  \item \texttt{'mb\char`_matchingscale'}, \texttt{'mc\char`_matchingscale'} -- set the scales (in GeV) where WET is matched onto WET-4 and WET-4 onto WET-3. Default to 4.2 and 1.3, respectively.
\end{itemize}

\subsection{Interfacing with other codes}\label{sec:interface}

Since \wilson\ builds on the Wilson coefficient exchange format WCxf, it is straightforward to import and export from and to programs supporting this standard. While the import has already been discussed above, the export can simply leverage the methods provided by the \texttt{wcxf} Python package, e.g.
\begin{lstlisting}[language=iPython]
wc = mywilson.match_run(scale=100, eft='WET', basis='JMS')
with open('my_wcxf_output.json', 'w') as f:
    wc.dump(f)
\end{lstlisting}

An even simpler data exchange is possible for codes written in Python themselves. In particular, the \texttt{flavio} package \cite{Straub:2018kue}, that can compute predictions for a plethora of observables in quark and lepton flavour physics, directly makes use of the \wilson\ package for the RG evolution, matching, and translation, starting from version v0.28. Functions that accept new physics Wilson coefficient values can be directly provided with a \texttt{Wilson} instance. This also allows to compute observables in terms of SMEFT Wilson coefficients. For example,
\begin{lstlisting}[language=iPython]
from wilson import Wilson
import flavio
mywilson = Wilson({'lq3_3333': 1e-6},
                   scale=1e3, eft='SMEFT', basis='Warsaw')
flavio.np_prediction('Rtaul(B->D*lnu)', mywilson)
\end{lstlisting}
computes the observable $R_{D^*}$ given a value of $1/\text{TeV}^2$
for the Wilson coefficient of the SMEFT operator
\begin{equation}
\left[O_{lq}^{(3)}\right]_{3333}  = \left( \bar \ell_3 \gamma_\mu \tau^I \ell_3 \right) \left( \bar q_3 \gamma^\mu \tau^I q_3 \right)\,,
\end{equation}
at the scale 1~TeV. The SMEFT running, matching, WET running, and conversion to the flavio basis used in the calculation of the observable is done behind the curtains by \wilson.

\section{Example}\label{sec:phys}

An interesting example where SMEFT RG effects lead to important constraints
on NP scenarios was discussed in refs.~\cite{Feruglio:2016gvd,Feruglio:2017rjo,Cornella:2018tfd}. It investigates scenarios attempting to simultaneously explain the deviations from lepton flavour universality observed in $b\to s\ell^+\ell^-$ transitions (with $\ell=e$ vs.\ $\mu$ ) and $b\to c\tau\nu$ transitions (with $\ell=\tau$ vs.\
$e$ or $\mu$)\cite{Aaij:2014ora,Aaij:2017vbb,Lees:2013uzd,Huschle:2015rga,Aaij:2015yra,Abdesselam:2016xqt},
\begin{align}
R_{K^{(*)}}^{\mu/e} &= \frac{\mathcal{B}(B \to K^{(*)} \mu^+ \mu^-)}{\mathcal{B}(B \to K^{(*)} e^+ e^-)}
\,,
&
R_{D^{(*)}}^{\tau/\ell} &= \frac{\mathcal{B}(B \to D^{(*)} \tau \bar \nu)_{exp}/\mathcal{B}(B \to D^{(*)} \tau \bar \nu)_{SM}}{\mathcal{B}(B \to D^{(*)} \ell \bar \nu)_{exp}/\mathcal{B}(B \to D^{(*)} \ell \bar \nu)_{SM}}
\,.
\end{align}
Since NP effects in semi-leptonic four-fermion operators with all left-handed fields are well known to fit the low-energy flavour data~\cite{Altmannshofer:2017yso, Capdevila:2017bsm, Alok:2017sui, Geng:2017svp, Ciuchini:2017mik}, it is interesting to consider NP models coupling dominantly to the third generation of left-handed quarks and leptons, such that the $b\to c\tau\nu$ transition, generated at tree level in the SM, receives sizable NP contributions, while the $b\to s\mu\mu$ transition is suppressed by flavour mixing angles that are assumed to be small~\cite{Calibbi:2015kma,Bhattacharya:2016mcc}. It was then shown that strong constraints arise on the simultaneous explanation of charged and neutral current anomalies from lepton flavour non-universality induced in leptonic tau decays, from lepton flavour violating tau decays, and from $Z$ pole observables.

The scenario considered in \cite{Feruglio:2017rjo} corresponds to the presence of the operators $[O_{lq}^{(1,3)}]_{3333}$ at a scale $\Lambda$ in
some weak basis that is related to the mass basis by small mixing angles.
Choosing a definite weak basis, namely the one conventionally used for the Warsaw basis in WCxf \cite{Aebischer:2017ugx}, where the down-type quark and charged lepton masses are diagonal, the following Wilson coefficients are present at the scale $\Lambda$:
\begin{align}
\left[C_{lq}^{(1)}\right]_{ijkl}  &= \lambda^{\ell}_{ij} \lambda^{q}_{kl} \,C_1
\,,
&
\left[C_{lq}^{(3)}\right]_{ijkl}  &= \lambda^{\ell}_{ij} \lambda^{q}_{kl} \,C_3
\,.
\end{align}
Assuming without loss of generality $\lambda^q_{33}=\lambda^{\ell}_{33}=1$ and adopting the simplified scenario where $\lambda^{q,\ell}_{22} = (\lambda^{q,\ell}_{23})^2$, a scenario can be initialized in \wilson\ as a function
of the parameters
$\mathtt{C1}=C_1$, $\mathtt{C3}=C_3$,
$\mathtt{lq_{-}  {23}}=\lambda^q_{23}$,
$\mathtt{ll_{-}} {{23}}=\lambda^{\ell}_{23}$
and $\mathtt{Lambda}=\Lambda$
as
\begin{lstlisting}[language=iPython]
from wilson import Wilson
ll_33 = ...
...
w = Wilson({'lq3_3333': ll_33 * lq_33 * C3,
            'lq1_3333': ll_33 * lq_33 * C1,
            'lq3_2223': ll_22 * lq_23 * C3, ...},
            scale=Lambda, eft='SMEFT', basis='Warsaw')
\end{lstlisting}
where the parameter variables have been set to numerical values\footnote{Note: \wilson\ does not accept symbolic inputs.}.
This \texttt{Wilson} instance can now be used to compute predictions for the relevant constraints using \texttt{flavio}, as discussed in section~\ref{sec:interface}:
\begin{lstlisting}[language=iPython]
from flavio import np_prediction
np_prediction('<Rmue>(B+->Kll)', w, 1, 6)  # R_K from 1-6 GeV^2
np_prediction('Rtaul(B->D*lnu)', w)        # R_D*(tau/l)
np_prediction('Rmue(B->D*lnu)',w)          # R_D*(mu/e)
np_prediction('BR(B+->Knunu)', w)          # B -> K nu nu
np_prediction('BR(tau->mumumu)', w)        # tau -> 3 mu
np_prediction('BR(tau->rhomu)', w)         # tau -> rho mu
np_prediction('BR(tau->enunu)', w)         # tau -> e nu nu
\end{lstlisting}

\begin{figure}
\includegraphics[width=0.5\textwidth]{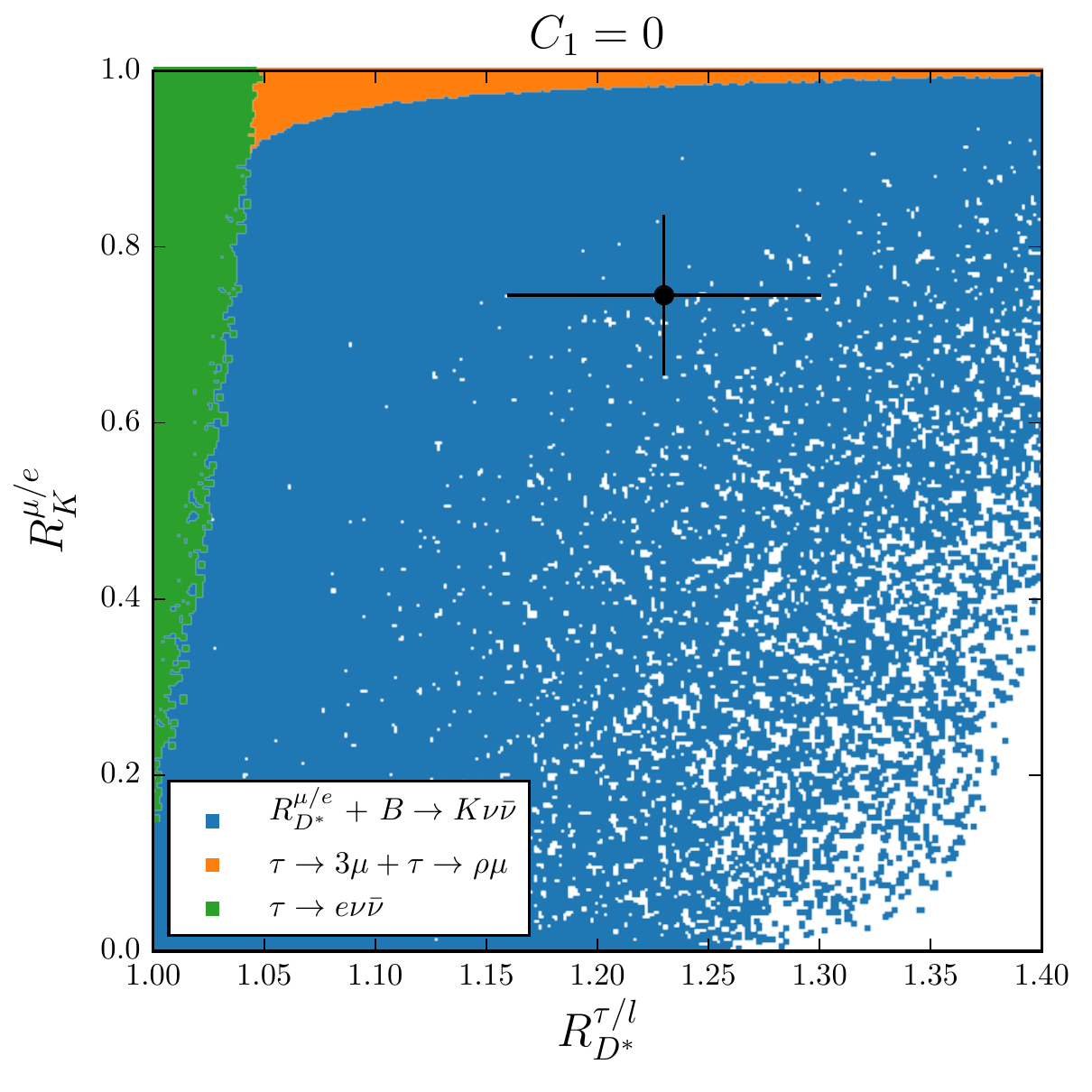}%
\includegraphics[width=0.5\textwidth]{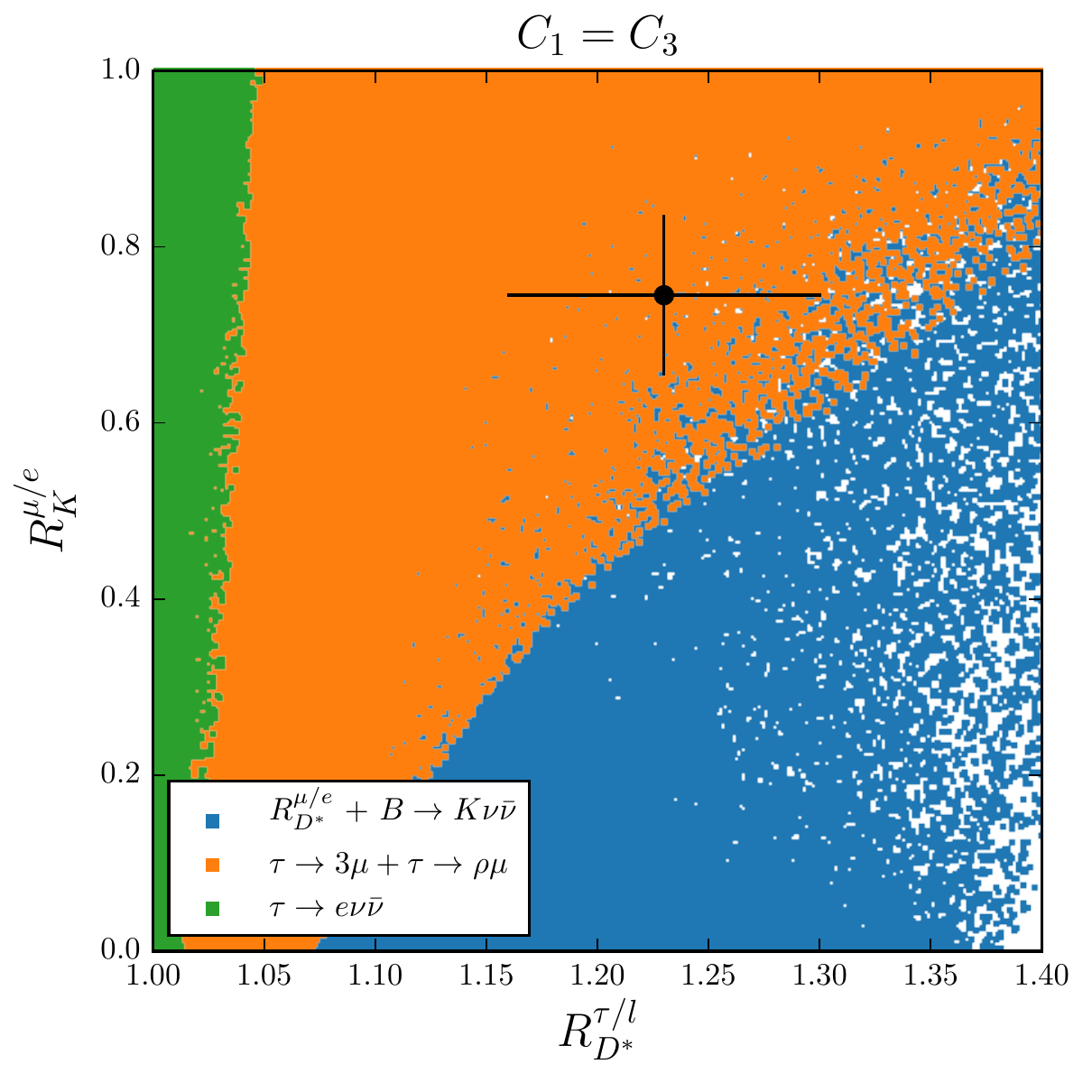}
\caption{Constraints on simultaneous solutions to $B$ anomalies through left-handed currents dominantly coupling to the third generation, reproducing fig.~5 of \cite{Feruglio:2017rjo}.}
\label{fig:padova}
\end{figure}

Using this procedure, in fig.~\ref{fig:padova} we have reproduced the result of refs.~\cite{Feruglio:2016gvd,Feruglio:2017rjo}, where the four free parameters are scanned as: $\lambda_{23}^q \in [-0.05,0]$,  $\lambda_{23}^{\ell} \in [-0.5,0.5], C_{1,3} \in [-4,0]$, and the scale $\Lambda$ is set at 1 TeV.
This shows that a simultaneous explanation of the charged and neutral current anomalies is disfavoured in this simplified scenario.

\section{Summary}

We have presented \wilson, a Python package for the RG evolution, matching, and basis translation of Wilson coefficients beyond the SM.
Starting from numerical values of Wilson coefficients at a high scale $\Lambda$, it automatically performs the necessary steps to return the Wilson coefficients at low energies relevant for precision measurements probing physics beyond the SM.
Built on the Wilson coefficient exchange format (WCxf), \wilson\ can be easily linked with a number of public codes, e.g.\ to directly compute the predictions for low-energy observables, as demonstrated in section~\ref{sec:phys}.

While \wilson\ is currently limited to one-loop RG evolution in SMEFT and WET and to tree-level matching, the structure of the code is general enough to be generalized to higher loop orders in the running and to loop-level matching (which is partially known, see e.g.\ \cite{Aebischer:2015fzz}) in the future. It has already been used in several NP analyses in the context of $B$ anomalies \cite{Kumar:2018kmr} and $\varepsilon/'\varepsilon$ \cite{Aebischer:2018rrz,Aebischer:2018quc,Aebischer:2018csl}. Being an open source project with a permissive license\footnote{\wilson\ is released under the MIT license.}, contributions from the community are welcome via the public code repository \cite{wilson-web}. Further information to \wilson\ can be found on the \wilson\ web page \url{https://wilson-eft.github.io/}.

\section*{Acknowledgements}

\noindent
The work of D.\,S. and J.\,A. is supported by the DFG cluster of excellence ``Origin and Structure of the Universe''. We thank Xuanyou Pan and Matthias Schöffel for important technical support in the development phase. J.\,K. thanks Michael Paraskevas for discussions.

\bibliographystyle{spphys}
\bibliography{bibliography}

\begin{thebibliography}{10}
\providecommand{\url}[1]{{#1}}
\providecommand{\urlprefix}{URL }
\expandafter\ifx\csname urlstyle\endcsname\relax
  \providecommand{\doi}[1]{DOI \discretionary{}{}{}#1}\else
  \providecommand{\doi}{DOI \discretionary{}{}{}\begingroup
  \urlstyle{rm}\Url}\fi

\bibitem{Weinberg:1967tq}
S.~Weinberg, Phys. Rev. Lett. \textbf{19}, 1264 (1967).
\newblock \doi{10.1103/PhysRevLett.19.1264}

\bibitem{Glashow:1961tr}
S.L. Glashow, Nucl. Phys. \textbf{22}, 579 (1961).
\newblock \doi{10.1016/0029-5582(61)90469-2}

\bibitem{Salam:1968rm}
A.~Salam, Conf. Proc. \textbf{C680519}, 367 (1968)

\bibitem{Buchmuller:1985jz}
W.~Buchmuller, D.~Wyler, Nucl. Phys. \textbf{B268}, 621 (1986).
\newblock \doi{10.1016/0550-3213(86)90262-2}

\bibitem{Grzadkowski:2010es}
B.~Grzadkowski, M.~Iskrzynski, M.~Misiak, J.~Rosiek, JHEP \textbf{10}, 085
  (2010).
\newblock \doi{10.1007/JHEP10(2010)085}

\bibitem{Appelquist:1974tg}
T.~Appelquist, J.~Carazzone, Phys. Rev. \textbf{D11}, 2856 (1975).
\newblock \doi{10.1103/PhysRevD.11.2856}

\bibitem{Wilson:1993dy}
K.G. Wilson, Rev. Mod. Phys. \textbf{55}, 583 (1983).
\newblock \doi{10.1103/RevModPhys.55.583}

\bibitem{Ellis:2018gqa}
J.~Ellis, C.W. Murphy, V.~Sanz, T.~You,   (2018)

\bibitem{Jenkins:2017jig}
E.E. Jenkins, A.V. Manohar, P.~Stoffer, JHEP \textbf{03}, 016 (2018).
\newblock \doi{10.1007/JHEP03(2018)016}

\bibitem{Aebischer:2017gaw}
J.~Aebischer, M.~Fael, C.~Greub, J.~Virto, JHEP \textbf{09}, 158 (2017).
\newblock \doi{10.1007/JHEP09(2017)158}

\bibitem{Jenkins:2017dyc}
E.E. Jenkins, A.V. Manohar, P.~Stoffer, JHEP \textbf{01}, 084 (2018).
\newblock \doi{10.1007/JHEP01(2018)084}

\bibitem{deFlorian:2016spz}
D.~de~Florian, et~al.,   (2016).
\newblock \doi{10.23731/CYRM-2017-002}

\bibitem{Weinberg:1980wa}
S.~Weinberg, Phys. Lett. \textbf{91B}, 51 (1980).
\newblock \doi{10.1016/0370-2693(80)90660-7}

\bibitem{Coleman:1969sm}
S.R. Coleman, J.~Wess, B.~Zumino, Phys. Rev. \textbf{177}, 2239 (1969).
\newblock \doi{10.1103/PhysRev.177.2239}

\bibitem{Callan:1969sn}
C.G. Callan, Jr., S.R. Coleman, J.~Wess, B.~Zumino, Phys. Rev. \textbf{177},
  2247 (1969).
\newblock \doi{10.1103/PhysRev.177.2247}

\bibitem{Alonso:2013hga}
R.~Alonso, E.E. Jenkins, A.V. Manohar, M.~Trott, JHEP \textbf{04}, 159 (2014).
\newblock \doi{10.1007/JHEP04(2014)159}

\bibitem{Straub:2018kue}
D.M. Straub.
\newblock {flavio: a Python package for flavour and precision phenomenology in
  the Standard Model and beyond} (2018).
\newblock \urlprefix\url{https://flav-io.github.io}

\bibitem{Jenkins:2013zja}
E.E. Jenkins, A.V. Manohar, M.~Trott, JHEP \textbf{10}, 087 (2013).
\newblock \doi{10.1007/JHEP10(2013)087}

\bibitem{Jenkins:2013wua}
E.E. Jenkins, A.V. Manohar, M.~Trott, JHEP \textbf{01}, 035 (2014).
\newblock \doi{10.1007/JHEP01(2014)035}

\bibitem{Aebischer:2015fzz}
J.~Aebischer, A.~Crivellin, M.~Fael, C.~Greub, JHEP \textbf{05}, 037 (2016).
\newblock \doi{10.1007/JHEP05(2016)037}

\bibitem{Aebischer:2017ugx}
J.~Aebischer, et~al., Comput. Phys. Commun. \textbf{232}, 71 (2018).
\newblock \doi{10.1016/j.cpc.2018.05.022}

\bibitem{Celis:2017hod}
A.~Celis, J.~Fuentes-Martin, A.~Vicente, J.~Virto, Eur. Phys. J.
  \textbf{C77}(6), 405 (2017).
\newblock \doi{10.1140/epjc/s10052-017-4967-6}

\bibitem{Herren:2017osy}
F.~Herren, M.~Steinhauser, Comput. Phys. Commun. \textbf{224}, 333 (2018).
\newblock \doi{10.1016/j.cpc.2017.11.014}

\bibitem{Kniehl:2016enc}
B.A. Kniehl, A.F. Pikelner, O.L. Veretin, Comput. Phys. Commun. \textbf{206},
  84 (2016).
\newblock \doi{10.1016/j.cpc.2016.04.017}

\bibitem{Bardeen:1978yd}
W.A. Bardeen, A.J. Buras, D.W. Duke, T.~Muta, Phys. Rev. \textbf{D18}, 3998
  (1978).
\newblock \doi{10.1103/PhysRevD.18.3998}

\bibitem{Dedes:2017zog}
A.~Dedes, W.~Materkowska, M.~Paraskevas, J.~Rosiek, K.~Suxho, JHEP \textbf{06},
  143 (2017).
\newblock \doi{10.1007/JHEP06(2017)143}

\bibitem{Gonzalez-Alonso:2016etj}
M.~González-Alonso, J.~Martin~Camalich, JHEP \textbf{12}, 052 (2016).
\newblock \doi{10.1007/JHEP12(2016)052}

\bibitem{Jung:2018lfu}
M.~Jung, D.M. Straub,   (2018)

\bibitem{Aebischer:2018iyb}
J.~Aebischer, J.~Kumar, P.~Stangl, D.M. Straub,   (2018)

\bibitem{Buchalla:1995vs}
G.~Buchalla, A.J. Buras, M.E. Lautenbacher, Rev. Mod. Phys. \textbf{68}, 1125
  (1996).
\newblock \doi{10.1103/RevModPhys.68.1125}

\bibitem{Buras:2011we}
A.J. Buras,   (2011)

\bibitem{wcxf-web}
{WCxf} web site.
\newblock \url{https://wcxf.github.io}

\bibitem{Feruglio:2016gvd}
F.~Feruglio, P.~Paradisi, A.~Pattori, Phys. Rev. Lett. \textbf{118}(1), 011801
  (2017).
\newblock \doi{10.1103/PhysRevLett.118.011801}

\bibitem{Feruglio:2017rjo}
F.~Feruglio, P.~Paradisi, A.~Pattori, JHEP \textbf{09}, 061 (2017).
\newblock \doi{10.1007/JHEP09(2017)061}

\bibitem{Cornella:2018tfd}
C.~Cornella, F.~Feruglio, P.~Paradisi, JHEP \textbf{11}, 012 (2018).
\newblock \doi{10.1007/JHEP11(2018)012}

\bibitem{Aaij:2014ora}
R.~Aaij, et~al., Phys. Rev. Lett. \textbf{113}, 151601 (2014).
\newblock \doi{10.1103/PhysRevLett.113.151601}

\bibitem{Aaij:2017vbb}
R.~Aaij, et~al., JHEP \textbf{08}, 055 (2017).
\newblock \doi{10.1007/JHEP08(2017)055}

\bibitem{Lees:2013uzd}
J.P. Lees, et~al., Phys. Rev. \textbf{D88}(7), 072012 (2013).
\newblock \doi{10.1103/PhysRevD.88.072012}

\bibitem{Huschle:2015rga}
M.~Huschle, et~al., Phys. Rev. \textbf{D92}(7), 072014 (2015).
\newblock \doi{10.1103/PhysRevD.92.072014}

\bibitem{Aaij:2015yra}
R.~Aaij, et~al., Phys. Rev. Lett. \textbf{115}(11), 111803 (2015).
\newblock \doi{10.1103/PhysRevLett.115.159901, 10.1103/PhysRevLett.115.111803}.
\newblock [Erratum: Phys. Rev. Lett.115,no.15,159901(2015)]

\bibitem{Abdesselam:2016xqt}
A.~Abdesselam, et~al.,   (2016)

\bibitem{Altmannshofer:2017yso}
W.~Altmannshofer, P.~Stangl, D.M. Straub, Phys. Rev. \textbf{D96}(5), 055008
  (2017).
\newblock \doi{10.1103/PhysRevD.96.055008}

\bibitem{Capdevila:2017bsm}
B.~Capdevila, A.~Crivellin, S.~Descotes-Genon, J.~Matias, J.~Virto, JHEP
  \textbf{01}, 093 (2018).
\newblock \doi{10.1007/JHEP01(2018)093}

\bibitem{Alok:2017sui}
A.K. Alok, B.~Bhattacharya, A.~Datta, D.~Kumar, J.~Kumar, D.~London, Phys. Rev.
  \textbf{D96}(9), 095009 (2017).
\newblock \doi{10.1103/PhysRevD.96.095009}

\bibitem{Geng:2017svp}
L.S. Geng, B.~Grinstein, S.~Jäger, J.~Martin~Camalich, X.L. Ren, R.X. Shi,
  Phys. Rev. \textbf{D96}(9), 093006 (2017).
\newblock \doi{10.1103/PhysRevD.96.093006}

\bibitem{Ciuchini:2017mik}
M.~Ciuchini, A.M. Coutinho, M.~Fedele, E.~Franco, A.~Paul, L.~Silvestrini,
  M.~Valli, Eur. Phys. J. \textbf{C77}(10), 688 (2017).
\newblock \doi{10.1140/epjc/s10052-017-5270-2}

\bibitem{Calibbi:2015kma}
L.~Calibbi, A.~Crivellin, T.~Ota, Phys. Rev. Lett. \textbf{115}, 181801 (2015).
\newblock \doi{10.1103/PhysRevLett.115.181801}

\bibitem{Bhattacharya:2016mcc}
B.~Bhattacharya, A.~Datta, J.P. Guévin, D.~London, R.~Watanabe, JHEP
  \textbf{01}, 015 (2017).
\newblock \doi{10.1007/JHEP01(2017)015}

\bibitem{Kumar:2018kmr}
J.~Kumar, D.~London, R.~Watanabe,   (2018)

\bibitem{Aebischer:2018rrz}
J.~Aebischer, A.J. Buras, J.M. Gérard,   (2018)

\bibitem{Aebischer:2018quc}
J.~Aebischer, C.~Bobeth, A.J. Buras, J.M. Gérard, D.M. Straub,   (2018)

\bibitem{Aebischer:2018csl}
J.~Aebischer, C.~Bobeth, A.J. Buras, D.M. Straub,   (2018)

\bibitem{wilson-web}
{wilson} github repository.
\newblock \url{https://github.com/wilson-eft/wilson}

\end{thebibliography}

\end{document}